\documentclass[aps,twocolumn,showpacs,twoside]{revtex4}

\usepackage{epsfig}
\usepackage{amsmath,amssymb,color}
\usepackage[english]{babel}

\parskip=\medskipamount

\newcommand{\eq}[1]{(\ref{#1})}
\newcommand{\fig}[1]{Fig.\ref{#1}}

\newcommand{\be}{\begin{equation}}
\newcommand{\ee}{\end{equation}}
\newcommand{\barr}{\begin{array}}
\newcommand{\earr}{\end{array}}
\newcommand{\beqn}{\begin{eqnarray}}
\newcommand{\eeqn}{\end{eqnarray}}
\newcommand{\bs}{\begin{subequations}}
\newcommand{\es}{\end{subequations}}

\def\runninghead#1#2{\pagestyle{myheadings}
\markboth{{\protect\it{\quad #1}}\hfill} {\hfill{\protect\it{#2\quad}}}}

\begin{document}

\runninghead{L.I. Nazarov, M.V. Tamm, V.A. Avetisov, S.K. Nechaev}{Statistical model of
intra-chromosome contact maps}

\title{Statistical model of intra-chromosome contact maps}

\author{L. Nazarov$^1$, M.V. Tamm$^{1,2}$, S.K. Nechaev$^{2,3,4}$, V.A. Avetisov$^{2,5}$}

\affiliation{$^1$Physics Department, M.V. Lomonosov Moscow State University, 119992 Moscow, Russia \\
$^2$Department of Applied Mathematics, Higher School of Economics, 101000 Moscow, Russia \\
$^3$Universit\'e Paris-Sud/CNRS, LPTMS, UMR8626, 91405 Orsay, France \\ $^4$P.N. Lebedev Physical Institute,
RAS, 119991 Moscow, Russia \\ $^5$N.N. Semenov Institute of Chemical Physics, RAS, 119991 Moscow, Russia}
\date{\today}

\begin{abstract}
The statistical properties of intra-chromosome maps obtained by a genome-wide chromosome
conformation capture method (Hi-C) are described in the framework of the hierarchical crumpling
model of heteropolymer chain with quenched disorder in the primary sequence. We conjecture that the
observed Hi-C maps are statistical averages over many different ways of hierarchical genome
folding, and show that the existence of quenched primary structure coupled with hierarchical
folding can induce the observed fine structure of intra-chromosome contact maps.
\end{abstract}


\maketitle

\begin{figure}[ht]
\epsfig{file=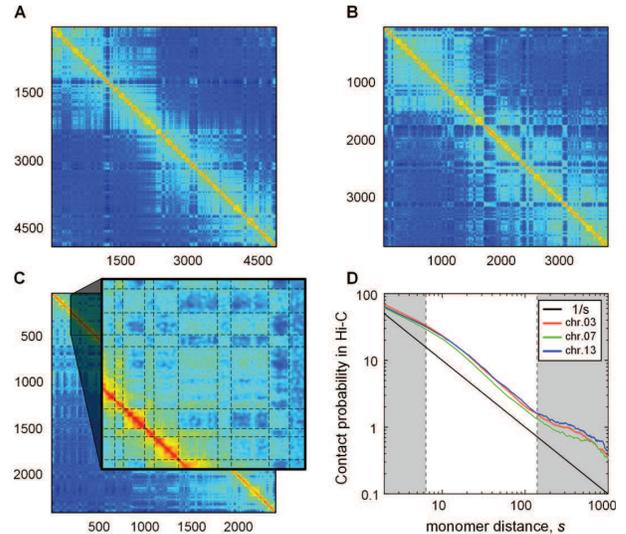,width=8cm} \caption{A-C: Typical samples of Hi-C maps (chromosomes 1 (A), 4 (B), and
13 (C), data used here provided to us by courtesy of M. Imakaev), the color encodes the contact probability
between genome fragments, each pixel corresponds to 2kb genome length; D: Contact probability decay in
doubly-logarithmic coordinates for the same Hi-C maps ($s$ is the genomic distance form the main diagonal of
a map) compared with $1/s$ power law.}
\label{fig:01}
\end{figure}

The analysis of DNA folding in human genome based on the genome-wide chromosome conformation capture method
\cite{3c,dekker} provides a comprehensive information on spatial contacts between genomic parts and imposes
essential restrictions on available 3D genome structures. The Hi-C maps obtained in the experiments on
various organisms and tissues \cite{dekker,dixon,drosoph,mouse,sofueva,dekkerreview,bacteria} (some examples
are shown in the \fig{fig:01}A-C) define the average contact probability, ${\cal P}(s)$ between two
particular parts of genome separated by the (genomic) distance $s$ along a chain. These maps display very
rich structure in a broad interval of scales, where ${\cal P}(s)\sim 1/s$ (see the \fig{fig:01}D) in
consistency with the so-called fractal (or "crumpled") globule \cite{gns} model of polymer chain packing. The
latter is characterized by two main features: (i) it has almost no knots, and (ii) it is self-similar in a
wide range of scales, forming a fractal space-filling structure. Both these properties are essential for
genome folding: fractal organization makes genome tightly packed at all scales, while the lack of knots
ensures easy and independent opening and closing of genomic domains, necessary for transcription
\cite{gr-rab,mirny2}. In 3D this packing results in a \emph{space-filling} conformation of a chain with
fractal dimension $D_f=D=3$.

The Hi-C contact probability, $P_{i,j}$, between two genomic units $i$ and $j$ depends in a complicated way
on a combination of structural and energetic factors. However, in a fractal globule with $D_f=3$ the average
contact probability, ${\cal P}(s) = (N-s)^{-1} \sum_{i=0}^{N-s} P_{i,i+s}$, between two units separated by
the genomic distance $s=|i-j|$ should decay as ${\cal P}(s) \sim s^{-1}$. This probability can be naively
estimated as ${\cal P}(s)\sim 1/R^D(s)$, where $R(s)$ is the typical size occupied by the $s$--fragment. For
curves with the fractal dimension $D_f$, one has $R(s) \sim s^{1/D_f}$, hence, ${\cal P}(s)\sim s^{-D/D_f}$,
and for $D_f=D=3$ one gets ${\cal P}(s)\sim 1/s$.

Apart from the averaged contact probability decay, one notes in Hi-C maps a chess-board-like intermittency in
color intensity \cite{julien}, elements of a hierarchical structure on small scales \cite{dekkerreview} (see
the inset in the \fig{fig:01}C) and chromosome compartmentalization on the large scale. Below we propose a
statistical toy model, which allows to reconstruct the $1/s$ decay of the contact probability, and the
typical fine structure of the experimentally observed Hi-C maps. Our model, on one hand, is based on sound
physical principles dealing with the fractal globule formation, and on the other hand, is kept as simple as
possible.

\begin{figure}[ht]
\epsfig{file=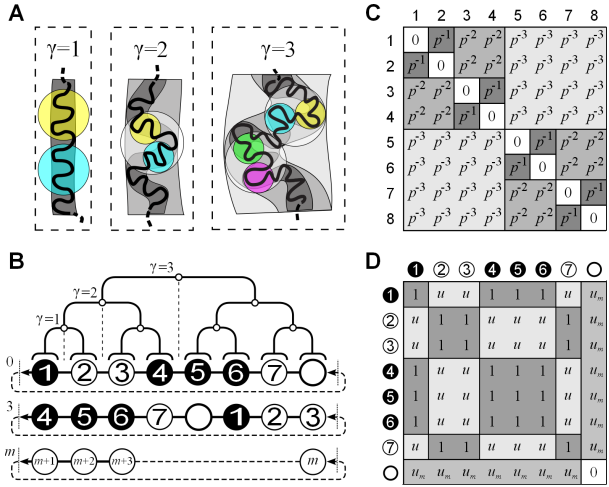, width=8cm} \caption{A: Schematic representation of a hierarchical folding; B:
Tree-like organization of crumples in a hierarchical folding; different translations are enumerated by the
parameter $m_f$; C: Block-hierarchical Parisi matrix of contact probabilities corresponding to a particular
folding scheme with $m_f=0$; D: Matrix of contact energies for a given primary structure.}
\label{fig:02}
\end{figure}

For the structural contribution to Hi-C probabilities, we assume a fractal globule model of genome
packing \cite{gr-rab}. The formation of a fractal globule (see \cite{gns}) is a hierarchical
process. In a collapsing polymer chain, there exists a certain chain-dependent critical length,
$g^*$, constituting the 1st-level fold (see the \fig{fig:02}A). As soon as 1st-level folds are
formed, the ``folds of folds'' of the second level start forming, etc., producing a hierarchy of
crumples, the process ends when all $g^*$-link folds are collected in a single (largest) fold.

In a hierarchically folded macromolecule each unit is characterized by a set of indices specifying
to which particular 1st-level fold (embedded into particular 2nd-level fold, \emph{etc}) this unit
belongs. This hierarchy of folds  can be naturally visualized as a Cayley tree of crumples, the
indexing is encoded by a path on the tree. The boundary nodes (leaves) of the tree constitute a
space of states for the units, and each subtree corresponds to a particular crumple. The number of
downward tree branches, $p$, defines the number of $\gamma$-level folds embedded in one fold of the
next level, $\gamma+1$. For simplicity, we assume here this embedding to be level-independent, also
for the length of an elementary unit of the chain, $l$, we suppose $l p = g^*$, meaning that the
smallest fold consists of $p$ elementary units (these units are, of course, much larger than single
nucleotides). Thus, each $\gamma$-level crumple consist of $p^\gamma$ units ($p=2$ in
\fig{fig:02}). The vertices at the tree boundary are enumerated ``along the chain'': we essentially
combine here the hierarchical (so-called "ultrametric") geometry for folding and the linear
geometry for underlying chain. Certainly, the assumed regularity of the hierarchy is an
oversimplification typical for a toy model and possible generalizations of the model are discussed
below. However, we believe that a regular tree model is a much more realistic 1st approximation to
the description of the hierarchical crumpling of a DNA chain into a fractal globule, than any sort
of random distribution without any reference to the hierarchy of crumples.

To account for the energetic contribution to the contact probability, we assume the elementary
units of a genome to be different from each other. The chessboard intermittency in the contact
probabilities typical for Hi-C maps (see \fig{fig:01}) suggests the existence of at least two
distinct types of units which we denote A and B (for the discussion of possible biological meaning
of this separation into different chromatin types see, e.g., \cite{julien}). The chess-board
intermittency of darker and lighter regions is modelled by an Ising-type energy cost, $E_{ij}$,
associated with the spatial contact between two units $i$ and $j$:
\be
E_{ij} = \left\{\begin{array}{ll} -1, & \mbox{if $i$ and $j$ are of same type} \medskip \\ -u, &
\mbox{otherwise}\end{array} \right.
\label{eq:1}
\ee
where $0\leq u \leq 1$ is the ratio of the energies of favorable and unfavorable contacts
(henceforth we use the energy of a favorable contact as the energy unit). Remind that the
elementary units A and B represent rather long chromatin fragments, consistent with the resolution
of the Hi-C method, which is of order of kilobases.

The probability $P_{i,j}$ of a contact between $i$th and $j$th units of a hierarchically folded
heteropolymer chain should depend on the generation, $\gamma_{ij}$, of the minimal common fold both
these units belong to, and on the species of the two units. We define the corresponding statistical
weight of the $ij$ contact as
\be
w_{ij}(\gamma)=e^{-\beta E_{ij}}P_{i,j}^{\rm str} + (1-P_{ij}^{\rm str}),
\label{eq:2}
\ee
where $E_{ij}$ is the contact energy \eq{eq:1} between units $i$ and $j$, $P_{i,j}^{\rm str}$ is an
\emph{a priori} (structural) contact probability between two units imposed by the hierarchy of
folds, $\beta$ is the inverse temperature, and the non-contact energy is assumed to be 0 by
definition. The structural probability is a function of $\gamma_{ij}$, and if the folding is
space-filling, then
\be
P_{ij}^{\rm str} \sim V^{-1}(\gamma_{ij}) \sim p^{-\gamma_{ij}},
\label{eq:2a}
\ee
where $V(\gamma_{ij})$ is the volume of the fold, which for the space-filling folding is
proportional to the number of elementary units in it. The proportionality coefficient in \eq{eq:2a}
can be absorbed into the definition of $E_{ij}$ so, without the loss of generality, $P_{ij}^{\rm
str}=p^{-\gamma_{ij}}$. In \fig{fig:02}C,D we depict binary matrix $E$ with elements $E_{i,j}$ and
a Parisi-type matrix $P$ with elements $P_{ij}^{\rm str}$ for some particular chain with a quenched
monomer sequence.

Although the hierarchical structure of folds emerges naturally within the fractal globule concept, its
presence in the experimental Hi-C maps is not always very evident. We explain that by suggesting that
\emph{smearing of the hierarchical block-diagonal structure occurs in experimental Hi-C maps due to the
non-uniqueness of DNA folding.}

Indeed, if the hierarchies of crumples are arranged differently in different folding realizations,
then the distinct block-diagonal structure, typical for each realization, is smeared out in the
ensemble averaging, while the $1/s$-decay of an average contact probability still holds. In what
follows we demonstrate that combining the hierarchical hteropolymer folding with the averaging over
different foldings, we can reproduce the typical behavior of experimentally observed Hi-C maps.

To introduce averaging over realizations we proceed as follows. Take a polymer chain with a given
sequence of units and consider all possible ways to fold it into hierarchical structures. Since in
our model the geometry of the tree of folds is fixed, and the chain fills all the folds
sequentially, there is only one possible way to alter the folding structure from one realization to
the other, that is, to change the position of the first elementary unit on the tree boundary -- see
\fig{fig:02}C. This change of the starting position induces a cyclic shift of all monomers along
the boundary: $i\to i+m\, ({\rm mod}\, p^{\gamma_{\max}}),\, i=1,2,...,N$), where $N$ is the length
of the chain. The only parameter defining a particular folding configuration is the shift $m$, in
\fig{fig:02}B the samples corresponding to $m=0$ and $m=3$ are shown.

Now, one can self-consistently define the weight of a particular hierarchical folding. Assume that
all contacts within a given folding are formed independently (i.e. all correlations in formation of
contacts are already encoded in the underlying tree structure). Then the total folding weight can
be written as a product of individual weights:
\be
W(m)=\prod_{i,j=1}^{p^{\gamma_{\rm max}}} w_{ij}(\gamma|m).
\label{eq:3b}
\ee
Here the weights $w_{ij}$ ($i,j=1..N<p^{\gamma_{\rm max}}$) are given by \eq{eq:2}, and one should
make an additional assumption about the weights of the contacts between the chain folds and ``outer
space'' (i.e., the chain parts surrounded by some other molecules), they are designated by open
circles in \fig{fig:02}b. Since the chromatin folding happens in a nuclei within a crowded
environment, one assumes that these open circles are effectively filled by the units of other
chromosomes. To account for this, we introduce a mean-field interaction between units of the chain
under consideration and the ``average'' units of other chains exactly as in Eq.\eq{eq:2} where
\be
E_{ij}=\left\{\begin{array}{ll} -q - u(1-q), & \mbox{if $i$ is of type A} \medskip \\
-(1-q) - qu, & \mbox{if $i$ is of type B}\end{array} \right.
\label{eq:3c}
\ee
and $q$ is an average fraction of monomers of type A.

The total partition function accounting for all folds, reads now
\be
Z=\sum_{m=0}^{p^{\gamma_{\max}}-1} W(m)
\label{eq:3d}
\ee
The probability for each pair of units, $i$ and $j$ is, as usual in equilibrium statistical mechanics
\cite{landafsh}
\be
P_{i,j}=\sum_{m=0}^{N-1}\frac{W(m)}{Z} \times \frac{e^{-\beta E_{ij}}P_{i,j}^{\rm
str}(m)}{e^{-\beta E_{ij}}P_{i,j}^{\rm str}(m) + (1-P_{i,j}^{\rm str}(m))},
\label{contact}
\ee
where the first term defines the thermodynamic probability of a particular hierarchical folding
realization, and the second term explicitly encounters the contact probability for this particular
folding. The values of $P_{i,j}$ given by \eq{contact} are the contact probabilities that should be
compared with the results of the Hi-C measurements for the intra-chromosome contact maps.

In the high-temperature limit, i.e. for $\beta=0$, we have $w_{ij}=1$ as it follows from \eq{eq:2},
and $P_{i,j}$ are just the averages of $P_{i,j}^{\rm str}$ over cyclic permutations of indices
along the Cayley tree boundary. Being averaged, all $P_{i,j}$ depend only on $s=|i-j|$, and in the
limit $N\gg 1$ they are ${\cal P}(s=|i-j|) \sim s^{-1}$. To reproduce this scaling, consider the
values of ${\cal P}(s)$ for $s = p^{m}, \; m=0,1,2...$. Let also $N$ be the power of $p$, i.e.
$N=p^M$. Then
\be
{\cal P}(p^m) =\frac{N}{N-1} \sum_{i=1}^{M-m} \frac{(1-p)}{p^i} \frac{1}{p^{(i+m)}} \simeq C p^{-m}
\label{eq:diag}
\ee
where $C=[p^2(1+p)]^{-1}$. The last equation is valid for $(M-m) \gg 1$. Since $p^{m}=s$, we get
${\cal P}(s)\sim s^{-1}$. This corresponds (at least in the high-temperature limit) to the
dependence typical for the real Hi-C data. This refines the naive arguments concerning
space-filling curves: any fold containing $\sim p^{\gamma}$ monomers is formed by a fragment of
length $s \sim p^{\gamma}$.

Let us summarize the main features of our model. Its input consists of a primary monomer sequence,
and three numerical parameters: (i) the number of subfolds in each fold of the hierarchy, $p$ (this
parameter, though being quantitatively important, does not influence the qualitative appearance of
the resulting structure of Hi-C maps), (ii) the comparative affinity of alternating AB contacts
versus AA/BB ones, $u$, and (iii) the inverse temperature, $\beta$, which essentially regulates the
uniqueness of folding. Indeed, for $\beta=0$ all foldings are equivalent and equally contribute to
the resulting probability, while for $\beta\to\infty$ the folding with the lowest energy give the
dominant contribution to the contact probabilities.

\begin{figure}[ht]
\epsfig{file=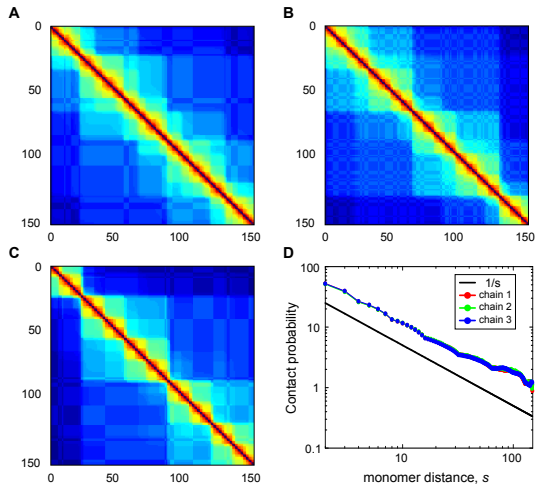,width=7cm} \caption{A-C: Dependencies of contact probability of the
hierarchical crumpled globule on the primary sequence ($N=150$) for different realizations of
primary sequences and fixed $u=0.25$, $\beta = 0.25$ and $p=2$; the average length of blocks is 10
for (A), 3.33 for (B) and 12.5 for (C); D: Averaged contact probability for sequences in A-C
compared to the $1/s$ plot.}
\label{fig:03}
\end{figure}

\begin{figure}[ht]
\epsfig{file=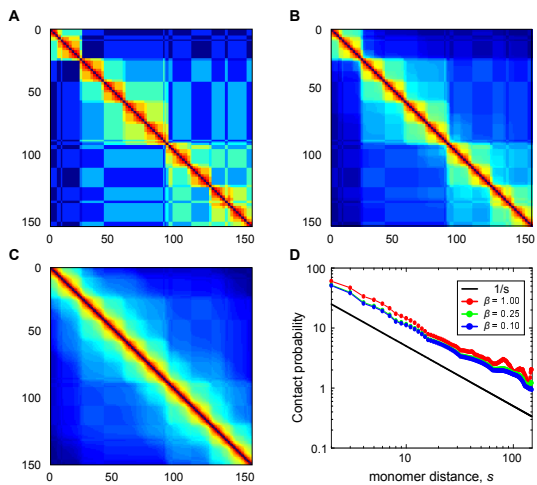,width=7cm} \caption{A-C: Dependencies of contact probability of the
hierarchical crumpled globule on the inverse temperature $\beta$ for fixed initial sequence of
$N=150$ monomers, with $u=0.25$ and $p=2$, A: $\beta=1$, B: $\beta=0.25$, C: $\beta=0.1$; D:
Comparison with $1/s$ plot.}
\label{fig:04}
\end{figure}

In \fig{fig:03} and \fig{fig:04} the examples of contact maps generated by our model are shown. To
demonstrate the influence of a monomer composition on the contact probability, we have generated
random Markovian heteropolymer sequences of $N=150$ monomers with varying average block length
(average lengths of blocks A and B are equal, thus the average fraction of units ``A'' is $q=0.5$).
The parameters $u=0.25$, $\beta=0.25$, and $p=2$ are the same for all three sequences. One sees
that the resulting contact map is essentially sequence-dependent, while the averaged contact
probability decay plotted in the \fig{fig:03}D as a function of genomic distance, $s$, still
follows the $1/s$ power law for all selected sequences.

In \fig{fig:04}A,B and C we show the influence of $\beta$ on the contact probability maps for a
fixed Markov chain of length $N=150$ with average block length 12.5 and $u=0.25$. One sees the
sequential degradation of the block-hierarchical structure with increasing temperature. The figure
\fig{fig:04}D demonstrates that the averaged contact probability is weakly temperature-dependent.

We see from \fig{fig:03} and \fig{fig:04} that our model indeed demonstrates a fine structure of
the contact probabilities reminiscent of that obtained in the experimental Hi-C maps. Note that the
existence of compartmentalization in the maps shown in the figures (i.e., large-scale block
structure) is due to the quenched disorder primary sequence. The configuration of large-scale
blocks is disorder-dependent (see figure \fig{fig:03}). In absence of any disorder all folding
configurations have equal weights, so the average contact probability decays gradually as $1/s$
with genomic distance $s$. Of course, the model presented above is a mere caricature of a real
situation: the number of possible folding ways in our model grows linearly with the chain length,
while it is bound to be exponential in real folding, when the hierarchical folding mechanism
accounts for intrinsic randomness. However, we believe that the main result concerning the
interplay of the uniqueness of folding and visibility of contact map fine structure will persist.
%will persist: i) the
%primary structure of heteropolymer chain together with particular way of hierarchical folding
%define the system energy; ii) if different foldings in the ensemble have essentially different
%energies, then the conformations with lowest energy dominate and the fine structure of the contact
%map is seen, while for degenerated conformations (with similar energies), the smearing of fine
%structure of contact maps occurs.

One possible way to generalize our model is as follows. Consider the hierarchy of folds, described by the
terminal points at the Cayley tree boundary (see the \fig{fig:02}) as being entirely separated from the chain
itself. In this case the hierarchy of folds plays a role of a specially prepared space of possible
microstates (conformations), in which the chain is embedded. In such a description transitions between
different folded conformations, i.e. transition between different terminal points at the Cayley tree
boundary, are described by the ``ultrametric'' random walk, where the transition weight form one state to the
other one is controlled by the highest barrier separating these two states \cite{ultram}. Then, the
coincidence of two folded configurations can be identified with the return probability of the ultrametric
random walk.

In this letter we have proposed a statistical model which reproduces the principal features of the
experimentally observed Hi-C maps. We have included into consideration the heteropolymer structure
equipped by the hypothesis of hierarchical DNA folding. We tried to avoid as much as possible the
specific ``biological'' details, sticking mainly to basic principles of statistical physics of
disordered systems. Such a description, being less informative for concrete biological systems,
allows us to conjecture the generic mechanism behind Hi-C map fine structure and could be
considered as a complimentary for the probabilistic refinement of Hi-C experiments developed
recently in \cite{ya-ta}.

We have assumed in our model that each single chromosome conformation is hierarchically folded with
distinct contact maps similar to hierarchical Parisi matrix shown in the \fig{fig:02}C. Since many
such conformations may have similar energies, in the resulting averaged structure of contacts the
hierarchical behavior could be smeared out. Besides, the largest compartments get smeared the
least, because the shift in the position of largest fold corresponds to the largest energy change.
We believe that the hierarchical compartmentalization of chromosomes into large domains, widely
observed in experiments is facilitated by a collective effect of many \emph{similar} microscopic
monomer-monomer interactions of heteropolymer primary structure folded as a fractal globule.

The authors are grateful to M. Imakaev, L. Mirny, A. Mironov and J. Mozziconacci for stimulating discussions.
This work was partially done during the visit of M.T. and L.N. to LPTMS, Universite Paris 6, the authors
appreciate the hospitality of their hosts as well as financial support from the IRSES project
FP7-PEOPLE-2010-IRSES 269139 DCP-PhysBio, which payed for this visit. M.T. and V.A. acknowledge the financial
support of the Higher School of Economics program for Basic Research.

\end{document}